\documentclass[aps,prb,twocolumn,groupedaddress]{revtex4-1}

\usepackage{color}
\usepackage{graphicx}
\DeclareGraphicsExtensions{.png}

\begin{document}


\title{Scanning gate imaging in confined geometries}


\author{R. Steinacher}
\email[]{richard.steinacher@phys.eth.ch}
\affiliation{Solid State Physics Laboratory, ETH Zurich, 8093 Zurich, Switzerland}

\author{A. A. Kozikov}
\affiliation{Solid State Physics Laboratory, ETH Zurich, 8093 Zurich, Switzerland}
\author{C. R{\"o}ssler}
\affiliation{Solid State Physics Laboratory, ETH Zurich, 8093 Zurich, Switzerland}
\author{C. Reichl}
\affiliation{Solid State Physics Laboratory, ETH Zurich, 8093 Zurich, Switzerland}
\author{W. Wegscheider}
\affiliation{Solid State Physics Laboratory, ETH Zurich, 8093 Zurich, Switzerland}
\author{K. Ensslin}
\affiliation{Solid State Physics Laboratory, ETH Zurich, 8093 Zurich, Switzerland}
\author{T. Ihn}
\affiliation{Solid State Physics Laboratory, ETH Zurich, 8093 Zurich, Switzerland}

\date{\today}

\begin{abstract}
This article reports on tunable electron backscattering investigated with the biased tip of a scanning force microscope. Using a channel defined by a pair of Schottky gates, the branched electron flow of ballistic electrons injected from a quantum point contact is guided by potentials of a tunable height well below the Fermi energy. The transition from injection into an open two-dimensional electron gas to a strongly confined channel exhibits three experimentally distinct regimes: one in which branches spread unrestrictedly, one in which branches are confined but the background conductance is affected very little, and one where the branches have disappeared and the conductance is strongly modified. Classical trajectory-based simulations explain these regimes at the microscopic level. These experiments allow us to understand under which conditions branches observed in scanning gate experiments do or do not reflect the flow of electrons.
\end{abstract}


\maketitle



Whoever has looked at twinkling starlight witnessed how atmospheric inhomogeneity randomly focuses light \cite{Lynch2001}. The inhomogeneity consists of small, random, but spatially correlated fluctuations of the refractive index. This rather common phenomenon is only one out of many examples, in which small random potentials generate regions of exceedingly high flow density, a bunching of particle trajectories, also known as caustics \cite{Berry1975}. In two-dimensional systems caustics form pairs of lines with an appearance like the branches of a tree. Topinka and coworkers were the first to observe such branches at the nano-scale in high quality two-dimensional electron gases \cite{topinka2001coherent}. In this system charged doping atoms randomly placed in a plane remote from the electron gas generate a smooth spatially correlated potential landscape. At liquid helium temperatures and below, the electrons emanating from a narrow constriction into this landscape exhibit branched flow \cite{heller2003branching}. Topinka observed it using the strongly repulsive electrostatic potential of a scanning tip to scatter channels of high flow density back through the constriction. Placing the tip along a caustic thereby measurably reduced the system's conductance. A spatially resolved image of the branch pattern was obtained by mapping the conductance as a function of tip-position.
The method, known as scanning gate microscopy \cite{Eriksson1996}, sparked hopes to observe other predicted phenomena of electron trajectories in nanostructured systems, such as wave function scars in ballistic stadiums. However, experimental success was very limited \cite{crook2003imaging, Burke2010,Ferry2011,Aoki2012}. Theoretically it is straight forward to calculate a backscattering pattern from known wave functions. The other way round, to extract the pattern of wave functions from an experimentally observed branch pattern is difficult to impossible. Furthermore, scanning gate measurements are by definition intrusive experiments and the conclusion one can draw about unperturbed wave functions are limited. Imaging and understanding the evolution of branches and caustics in geometries tunable between free and confined electron motion therefore remains an interesting experimental challenge. Here we tackle the problem using a sample-geometry simple enough to interpret the resulting conductance maps conclusively. Our experiments lead to striking general insights into the imaging mechanism and spatial resolution of the scanning gate technique in structures with tunable confinement.


\begin{figure}
	\centering
	\includegraphics[scale=1]{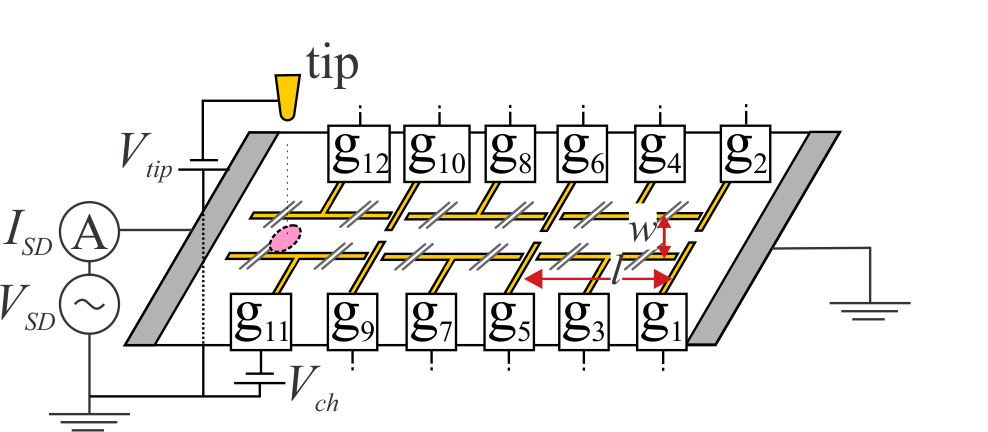}
	\caption{Schematic of gate geometry and measurement setup. Gates are labeled g$_1$ to g$_{12}$. The pink area indicates the movable tip-depleted region.}
	\label{fig1}
\end{figure}

We measured two similar samples (labeled A and B) that are based on a molecular-beam-epitaxy-grown GaAs/AlGaAs heterostructure (the same as in Refs. \onlinecite{kozikov2014locally, kozikov2013interference, steinacher2015scanning}) with a two-dimensional electron gas (2DEG) 120\,nm below the surface. The electron gas in sample~A has a density $n=1.4\times 10^{11}\,\mathrm{cm^{-2}}$ and a mobility $\mu = 9.3 \times 10^{6}\,\mathrm{cm^2}/\mathrm{Vs}$; for sample~B, $n=1.4\times 10^{11}\,\mathrm{cm^{-2}}$ and $\mu = 7.1 \times 10^{6}\,\mathrm{cm^2}/\mathrm{Vs}$, both at a temperature of 300\,mK. The structure was fabricated on a Hall bar of 200\,$\mu$m width and 2\,mm contact separation with Au/Ge/Ni ohmic contacts. Schottky gates (Ti/Au) defined by electron-beam lithography complete the device structure schematically shown in Fig.\,\ref{fig1}. It consists of three consecutive channels of width $w=1$\,$\mu$m (channel\,1 defined by gates g$_3$ and g$_4$, channel\,2 by g$_7$ and g$_8$, and channel\,3 by g$_{11}$ and g$_{12}$) and three quantum point contacts with a lithographic gap of 300\,nm (QPC\,1 defined by g$_1$ and g$_2$, QPC\,2 by g$_5$ and g$_6$, and QPC\,3 by g$_9$ and g$_{10}$). Neighboring quantum point contacts are separated by $l=15$\,$\mu$m along the channel axis. The Schottky gates deplete the 2DEG at voltages below $-0.35$\,V. 

The samples were mounted in a home-built scanning force microscope operated in a $^3$He-cryostat \cite{Ihn2004} with a base temperature of $300\,$mK. A phase-locked loop controlled the microscope's tuning fork sensor \cite{Rychen1999, Rychen2000}, to which a Pt/Ir wire was glued that had been sharpened in a focussed-ion-beam.  With a voltage $V_\mathrm{tip}=-6$\,V applied between tip and 2DEG, we raster-scanned the tip 60\,nm above the sample surface. At this voltage, the tip depleted the 2DEG within a disk of about 800\,nm diameter \cite{steinacher2015scanning}. 
While the tip scanned above the surface of the structure, we determined the two-terminal conductance by applying an alternating source--drain voltage $V_\mathrm{SD}=100\,\mu$V$_\mathrm{rms}$ to the Hall bar and measuring the alternating current $I_\mathrm{SD}$ with a home-built current--voltage converter and a commercial lock-in amplifier. In this way, we recorded maps $G(x,y)=I_\mathrm{SD}(x,y)/V_\mathrm{SD}$ of linear conductance vs. tip-position $(x,y)$.

\begin{figure}
	\centering
	\includegraphics[scale=1]{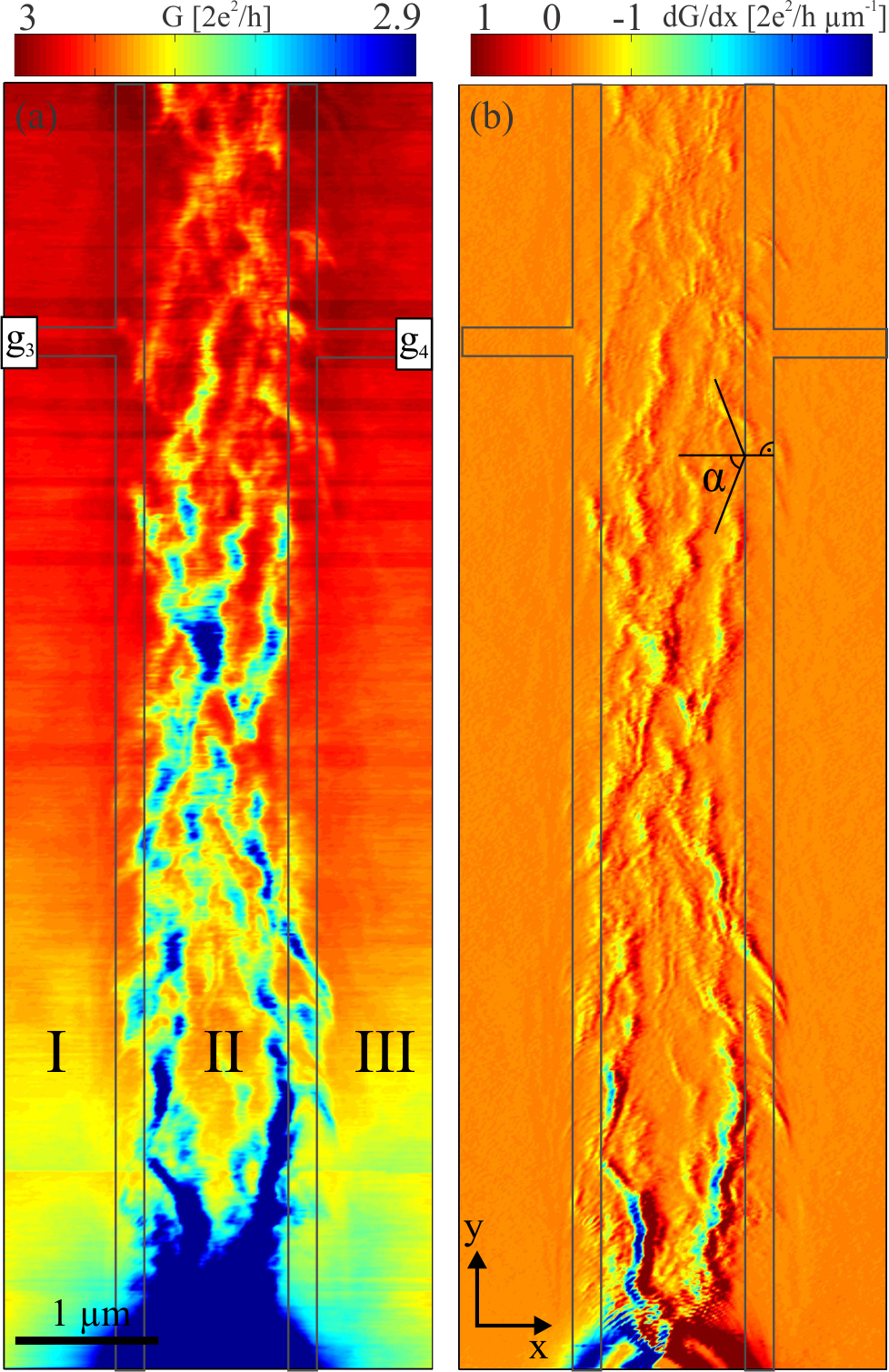} 
	\caption{(a) Conductance $G(x,y)$ of QPC\,1 on sample~A. Grey solid lines mark the outlines of the gates. Roman numbers label the channel region (II) and the two regions outside the channel (I, III). (b) Derivative $dG(x,y)/dx$ of the data in (a).}\label{fig2}
\end{figure}

In the following, we investigate branched electron flow in sample~A applying depleting voltages to the split gates g$_1$ and g$_2$, such that QPC\,1 has a quantized conductance of $3\times 2e^2/h$. All the other gates have 0\,V applied. The resulting scanning gate image in Fig. \ref{fig2}(a) has the outline of the gates superimposed with grey solid lines. In Fig. \ref{fig2}(b) the tip-position dependent variations of the conductance are emphasized by showing the horizontal derivative $dG(x,y)/dx$  of the conductance in Fig. \ref{fig2}(a). On a large scale, both images consist of three regions that we label I--III in Fig. \ref{fig2}(a), with region II being within the channel, and regions I and III outside. Most remarkably, the spatial conductance variations induced by the scanning tip are confined to the channel region II, whereas the conductance images are smooth in the outer regions I and III. This made us suspect that placing the tip-induced potential outside the channel does not lead to scattering of electrons back through the QPC.

The branching pattern of conductance variations observed within the channel region II are reminiscent of the branched electron flow that Topinka \cite{Topinka2000, topinka2001coherent} and later also Kozikov \cite{kozikov2013interference} found in scanning gate experiments, where a QPC tuned to a quantized value of the conductance injects electrons into a 2DEG reservoir. Heller explained the formation of these branches in Ref.~\onlinecite{heller2003branching} as a conspiracy between random focusing of electron flow in a weak long-range spatially fluctuating potential in the absence of the tip, and time reversal symmetry at zero magnetic field which leads to backscattering through the QPC of a selected subset of electron trajectories by the tip.

The finding in Fig.\,\ref{fig2}, that the branching pattern is confined to the channel region although we apply zero volts to the channel gates g$_3$ and g$_4$, suggests, that these gates induce a small potential barrier with height much less than the Fermi energy. The GaAs surface pins the Fermi energy roughly in the center of the band gap \cite{spicer1979new,spicer1979surface,colleoni2014origin}. Depositing the metallic gate on top of the surface nevertheless changes the surface potential by an amount small compared to the band gap, and thereby induces a small potential barrier in the 2DEG. In addition, strain fields originating from a difference of the thermal expansion coefficients of the gate metal and the semiconductor will induce a small potential in the 2DEG via deformation or piezoelectric coupling \cite{winkler1989landau, yagi1993oscillatory, davies1994theory, larkin1997theory, petticrew1998piezoelectric, kato1998two}. Therefore we interpret the confinement of the branching pattern of the conductance to be the result of a shallow potential barrier below the unbiased gate electrodes.

In regions I and III of Fig. \ref{fig2}(a), the smoothly varying change of the background conductance arises due to the long-range capacitive influence of the tip on the potential in QPC\,1. The closer the tip moves towards the constriction, the more will the saddle-point of the QPC potential be lifted towards the Fermi energy. As a consequence the conductance gradually reduces \cite{Topinka2000, topinka2001coherent, kozikov2013interference}.

\begin{figure*}[floatfix]
	\centering
	\includegraphics[scale=1]{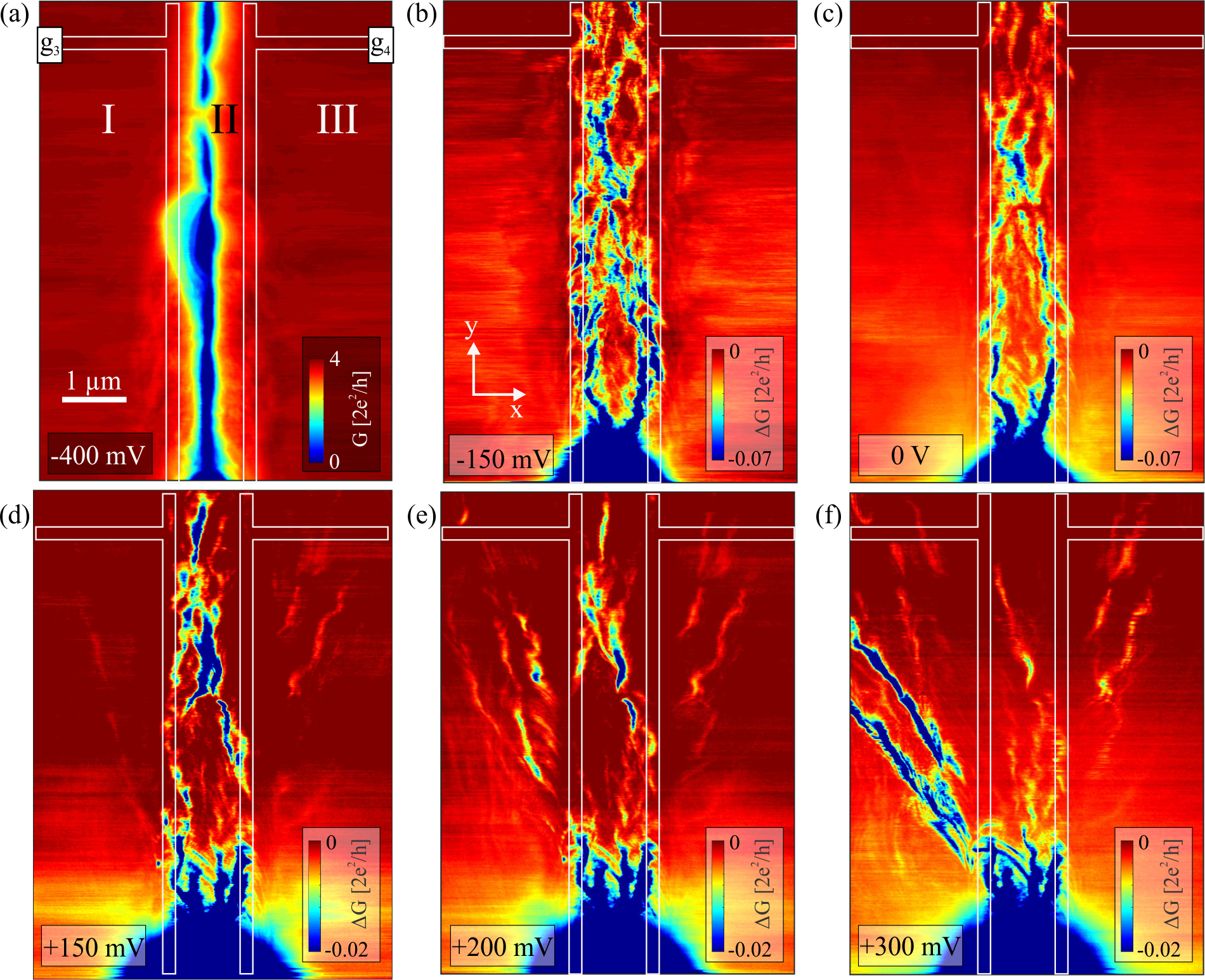} 
	\caption{Scanning gate images for different channel gate voltages as indicated in the images. (a) Conductance as a function of tip position for the QPC transmitting four spin-degenerate modes and $V_\mathrm{ch}=-400$\,mV to electrostatically define the channel. (b)-(f) Series of scans with different, non-depleting $V_\mathrm{ch}$ showing $\Delta G=G-4e^2/h$ in color. The QPC transmits 2 spin-degenerate modes in the absence of the tip.}\label{fig3}
\end{figure*}

In the next step we compensate the shallow potential barrier existing at zero applied voltage for electrons below gates g$_3$ and g$_4$ by the application of finite positive voltages. Figure \ref{fig3}(b)--(f) shows scanning gate images for a selection of gate-voltages between $-150$\,mV and $+300$\,mV, where the QPC supports two spin-degenerate modes. In this transmission mode the guiding of the branches can be seen best without making the density of branches inside the channel too high (for V$_\mathrm{ch}=0$\,V). Figure \ref{fig3}(c) reproduces Fig. \ref{fig2}(a) for convenience, whereas Fig. \ref{fig3}(d) was obtained with $+150$\,mV applied to the channel gates g$_3$ and g$_4$. Faint branches appear in this image in regions I and III, where no structure had been seen at zero applied gate voltage. The branches penetrate into these regions even more in Figs. \ref{fig3}(e) and (f), where even more positive voltages were applied to the channel gates. In these images, the presence of the channel gates cannot easily be guessed from the measured image without prior knowledge.

We see contrasting behavior in Figs. \ref{fig3}(b) and (c), where the channel gate voltage becomes increasingly negative. In Fig. \ref{fig3}(b) the conductance pattern reminds us of mesoscopic conductance fluctuations, but the underlying branch pattern seen in Fig. \ref{fig3}(c) still leaves its traces. The potential barrier below the channel-gate is still lower than the Fermi energy in Fig. \ref{fig3}(b).

Figure \ref{fig3}(a) shows a scan where the electron gas below the gates is completely depleted. In contrast to (b)--(f) we plot conductance $G$ rather than conductance change $\Delta G$. No matter what color-scale we choose for this image the branches have disappeared and given way to smooth variations of the conductance on a scale of $1\,\mu$m along the channel axis. Placing the tip in the center of the channel blocks transport completely, because electrons can not escape into regions I and III.

This series of measurements demonstrates that the branches of enhanced backscattering observed in scanning gate measurements near a QPC can be guided by intentionally patterned shallow potentials. This main experimental finding is natural given the notion that the observation of branches are a result of the shallow random potential landscape in the 2DEG. However, it opens new opportunities for controlling mesoscopic fluctuations in ultra-high mobility structures where artificial shallow potential landscapes steer the electron flow and thus dominate over the static potential fluctuations.

We may ask a number of questions at this point: how is it possible that the branching pattern is confined by a shallow potential barrier? Does the large and invasive tip-induced potential still image the electron flow in the absence of the tip as it does in an open 2DEG? How does the presence of the tip change the electron flow? What can we learn from our measurements for the general case of scanning gate measurements in confined geometries? Let us answer these questions by presenting additional analysis and by digging deeper into the microscopic physics of the experiment.

The branch-guiding property of a shallow potential may be seen as an effect of geometric electron optics. The potential barrier below the channel gates plays the role of a medium reflecting or transmitting incoming electron beams according to Snell's law. Electrons at the Fermi energy $E_\mathrm{F}$ impinging on the barrier of height $V$ at an angle (measured with respect to the barrier normal) larger than the critical angle $\alpha_\mathrm{c} = \arcsin\sqrt{1-V/E_\mathrm{F}}$ are totally reflected back into the channel with their momentum parallel to the channel axis conserved [see schematics in Fig. \ref{fig2}(b)]. In our structure we estimate $V/E_\mathrm{F}\approx 0.36$ at zero channel gate voltage \footnote{For this estimate we assume complete compensation of the barrier (flat band) at +200\,mV and pinch-off at -350\,mV. At 0\,V we therefore find $V/E_\mathrm{F}=200/(200+350)=0.36$. The precision of compensation-voltage is on the order of $\pm50$\,mV} giving $\alpha_\mathrm{c}=53^\circ$. Collimation of the electron beam due to QPC\,1 is expected to lead to an injection characteristic essentially cut off by a critical injection angle $\beta_\mathrm{c}$ (measured with respect to the channel axis) \cite{Beenakker1991, Crook2000a}. We may obtain an experimental estimate of the angular distribution of injected electrons from Fig. \ref{fig3}(c). In this figure we see that the observed backscattering branches outside the wire fan out at a maximum angle $\beta_\mathrm{c}\approx 45^\circ$. This maximum injection angle leads to a fraction of 85\% of electrons that would remain within the channel boundaries after the first reflection. The remaining 15\% are typically not visible because of the experimental limit in sensitivity.

\begin{figure*}
	\centering
	\includegraphics[scale=1]{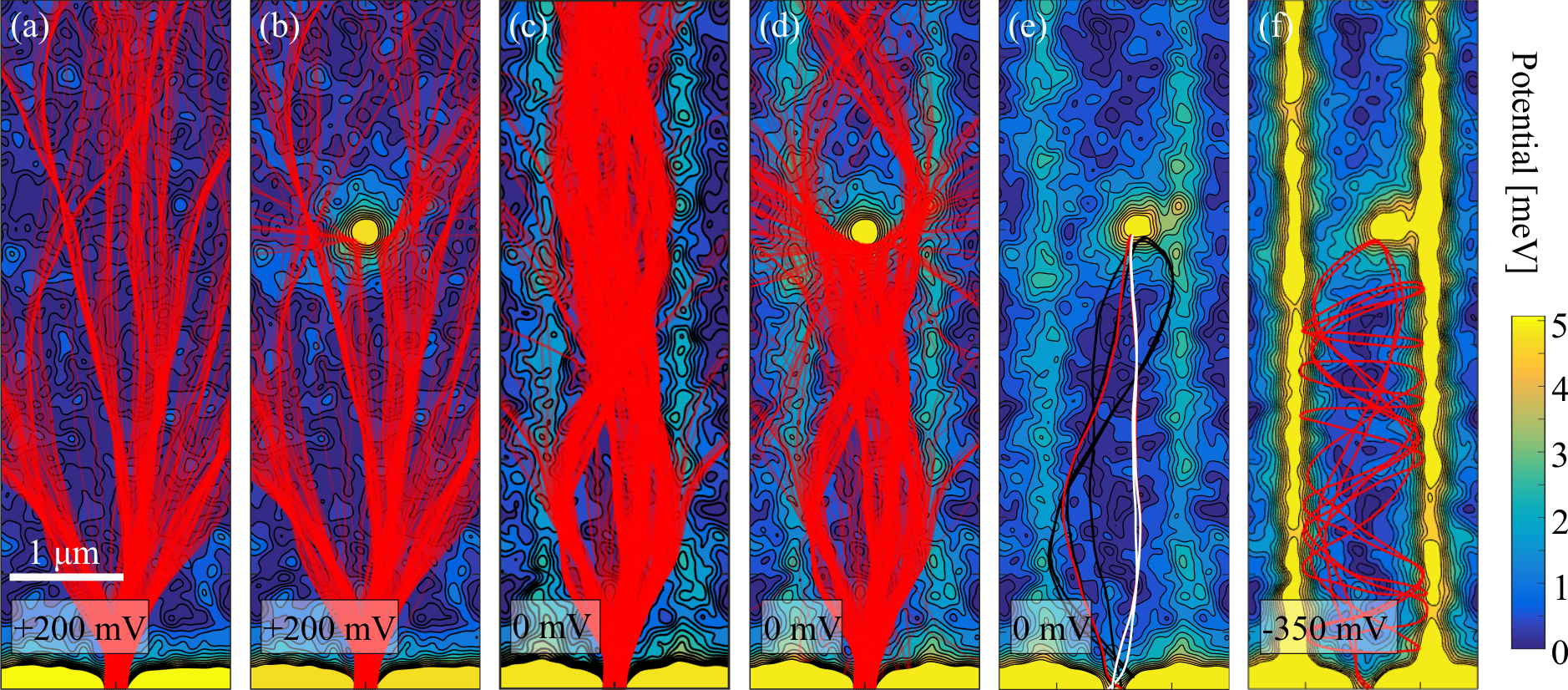} 
	\caption{Trajectories of electrons injected through the QPC (conductance $4e^2/h$) for different situations: (a) Only the QPC, no channel (V$_\mathrm{ch}=+200$\,mV), no tip. (b) QPC and tip, but no channel (V$_\mathrm{ch}=+200$\,mV). (c) QPC and shallow channel potential (V$_\mathrm{ch}=0$\,mV), no tip. (d) QPC and shallow channel potential (V$_\mathrm{ch}=0$\,mV) with tip. (e) Three distinct backscattered trajectories in QPC, tip, and shallow channel potential (V$_\mathrm{ch}=0$\,mV). (f) One backscattered trajectory with depleting QPC, channel (V$_\mathrm{ch}=-350$\,mV), and tip.}\label{fig4}
\end{figure*}

\begin{figure}
	\centering
	\includegraphics[scale=1]{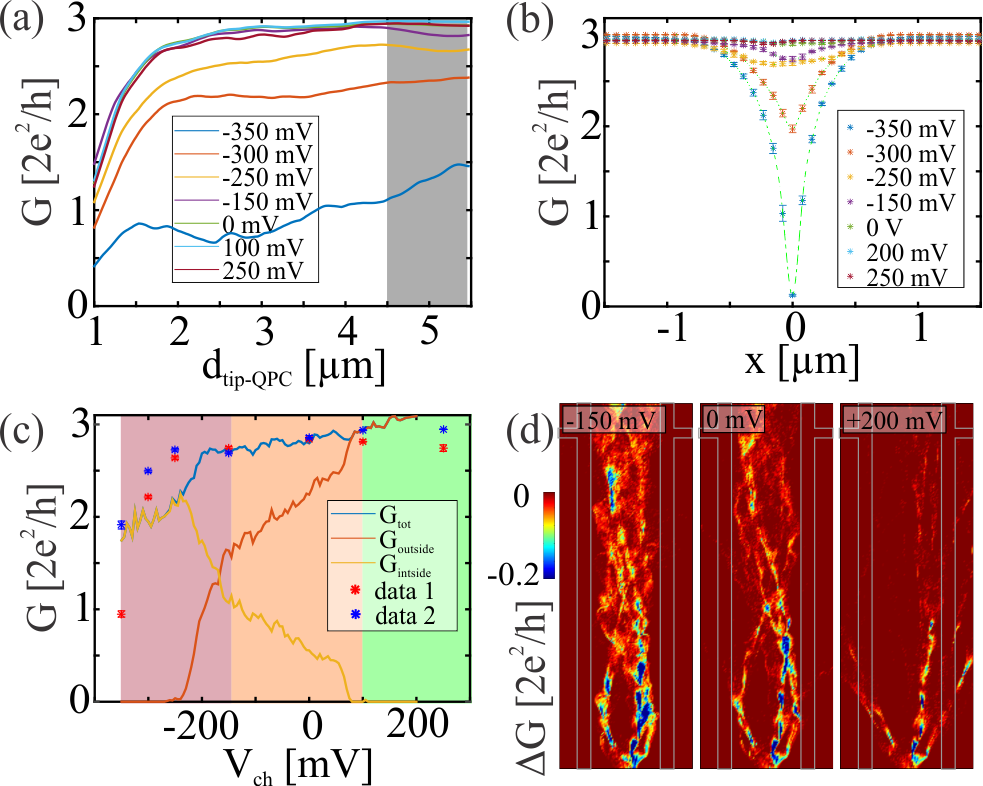} 
	\caption{(a) Smoothed conductance cuts along $y$, roughly $80$\,nm from the channel center of sample B for different channel gate voltages between complete pinch-off ($-350$\,mV) and flat band below channel gates ($+250$\,mV). (b) Smoothed conductance cuts along $x$ at $y=5\,\mu$m for a set of channel gate voltages. Smoothing extended over a range of 1\,$\mu$m [shaded region in (a)]. (c) Symbols: measured average conductance of the system with the tip in the center of the channel as a function of $V_\mathrm{ch}$. Solid lines: calculated transmission and reflection probabilities (see main text). (d) Simulated scanning gate images with a QPC conductance of $6e^2/h$ at different channel-gate voltages.}\label{fig5}
\end{figure}

Classical trajectory-based simulations of the conductance \cite{Beenakker1989,Baranger1991} help us to understand electron branching and the classical background resistance. We find that they qualitatively reproduce the experimental behavior [see Fig. \ref{fig5}(d)] and give further insight into the flow of electrons in the structure with and without the tip, and therefore on the imaging mechanism at work. We calculate trajectories by solving Newton's equations in two dimensions for a given potential landscape. Gate-induced potentials are modeled using the analytic expressions of Davies \cite{Davies1995}. A Lorentzian profile represents the tip-induced potential \cite{topinka2001coherent,Eriksson1996a}\footnote{Note that the tip-depletion size for the simulation is smaller than the one in the experiment to get similar scanning gate maps. Electrostatic screening of the tip-induced potential, finite resolution and signal-to-noise in the experiment naturally lead to less sensitivity compared to the simulation}. We introduce a random disorder potential landscape caused by ionized impurities in the doping plane which accounts for Thomas-Fermi screening \cite{Ando1982}, finite thickness of the 2DEG \cite{Fang1966, Ando1982}, and charge correlations in the doping plane \cite{Efros1990}. All the parameters of the disorder potential are given by the sample geometry and the electron density at zero gate-voltage, except for the correlation parameter which is used to tune the theoretical electron mobility to be the same as in the experiment. The conductance is a weighted sum of individual trajectory contributions.

In agreement with Topinka \cite{topinka2001coherent}, Heller \cite{heller2003branching}, and Metzger \cite{metzger2010universal, Metzger2010} our model reproduces the formation of caustics and branched electron flow in the absence of the scanning tip in an open 2DEG past a QPC [Fig. \ref{fig4}(a)]. Caustics are bundles of higher than average (in theory even diverging) trajectory-density caused by accidental lensing in the random disorder potential. Heller argued in Ref. \onlinecite{heller2003branching} that a hard-wall cylinder-shaped tip-potential would reflect branches with close to normal incidence back through the QPC on time-reversed paths. He explained in this way the superb spatial resolution of scanning gate microscopy which is much better than the diameter of the tip. A more realistic soft tip-potential modifies this description in two ways: First, the disorder potential distorts the Fermi-contour of the tip-induced potential changing the direction of normal incidence [see the distorted Fermi-contour in Fig. \ref{fig4}(b), (d), and (e)]. Second, the long-range tail of the tip-potential slightly diverts the branches present in the absence of the tip. Both effects lead to differences between the branched flow in the absence of the tip and the scanning-gate images. Less severe for the appearance of the scanning-gate image is the fact that the electron flow is strongly modified by the presence of the tip [c.f. Figs. \ref{fig4}(a) and (b)]: trajectories that are only diverted but not backscattered will not reduce the measured conductance. 

The trajectory model confirms that the branches are guided by shallow channel-potentials [see Fig. \ref{fig4}(c)]. Like in the open 2DEG the tip scatters most trajectories out of the channel but not back trough the QPC [Fig. \ref{fig4}(d)] and hence does not change the conductance. Nevertheless, an increasingly negative channel-gate voltage increases the proportion of trajectories scattered back through the QPC. In the open system only direct backscattering from the tip is possible [like the white trajectory in Fig. \ref{fig4}(e)], whereas the channel-potential gives rise to trajectories scattering once [see red curve in Fig. \ref{fig4}(e)] or multiple times [black curve in Fig. \ref{fig4}(e)] from the channel-gates before they are backscattered through the QPC. These trajectories include paths with normal incidence on the tip but also others that enclose a finite area. This implies that the ability to image branch-like classical trajectory-families is reduced the stronger the channel is confined. When the channel-gate voltage depletes the underlying electron gas most trajectories spend a long time in the cavity between the QPC and the tip, bouncing chaotically from wall to wall leaving the structure only accidentally through the QPC [see Fig. \ref{fig4}(f)] or the opening between the tip and the channel gate.

Since the spatial potential fluctuations are caused by the random distribution of ionized donors, the branching pattern differs from sample to sample, and even for different cool downs. To compare experiment and simulation not only qualitatively [see Fig. \ref{fig5}(d) for simulated scanning gate images] we therefore average the measured conductance, such that only the classical background remains. Accordingly, we compare these averaged quantities with simulations without a disorder potential.

To this end we analyzed a series of scanning gate images at QPC\,3 with different voltages applied to channel\,3 of sample~B. In Fig. \ref{fig5}(a) we plot the conductance of the system along a line parallel to the channel axis. We smoothen the cuts with a running average of $900$\,nm width to remove most of the modulations that are caused by branches from the background. It is not possible to remove them completely: variations on length scales larger than $900$\,nm remain.

In Fig. \ref{fig5}(a) the conductance increases with increasing tip--QPC distance reaching a constant value at $d_\mathrm{tip-QPC}>2\,\mu$m. The strong increase of the conductance within the first two microns is due to strong backscattering with the tip in close vicinity to the constriction as well as due to the gating effect. We obtain the curves shown in Fig. \ref{fig5}(b) by taking average saturation values [see grey-shaded region in (a)] of similar curves taken parallel to the channel-axis at different $x$-coordinates. The channel is seen as a pronounced dip in the conductance with a strength that increases with increasing confinement of the channel, as expected. Figure \ref{fig5}(c) shows the conductance values at $x=0$ of two channels (data 1 and data 2) on sample B plotted against the channel-voltage. The trajectory physics differs significantly in the regions colored differently. In the green-shaded region most electrons scattered off the tip are not backscattered through the QPC. Backscattering occurs preferentially along branches modulating the conductance at a level of a few percent. In the red region the electron flow is strongly channeled by the gate-induced potential in the absence of the tip. This effect leads to the observed guiding of branches. However, the channel-potential is still too low to prevent the majority of electrons scattering off the tip from propagating into regions I and III. This changes in the violet region where the electrons are increasingly kept inside the channel, enhancing their chance to scatter back through the QPC.

We compare the experimental behavior in Fig.\,\ref{fig5}(c) with the transmission and reflection probabilities calculated with the classical trajectory model. The transmission probability is split into two contributions, namely, the probability of being transmitted and leaving the structure through regions I or III (red denotes G$_\textrm{outside}$), and the probability of transmitting beyond the tip within the wire region II (yellow denotes G$_\textrm{inside}$). The model supports our interpretation by showing that in the green region almost all electrons are scattered out of region II whereas in the red region the electrons increasingly stay in the channel with growing wire potential. In the violet region, however, the number of electrons that make it past the tip starts to decrease, because the channels between tip and wire potential shrink.


In summary, we have studied branched electron flow in a wire geometry tunable between weak and strong confinement using scanning gate microscopy. A comprehensive understanding of the measured conductance was reached based on classical trajectories. Weak confinement guides the branches known from open two-dimensional electron gases. In contrast, stronger confinement generates a chaotic cavity with strongly enhanced backscattering.
Hand in hand with the change in trajectory dynamics the scanning gate technique gradually loses its spatial resolution for backscattered electron flow from the weakly to the strongly confined regime.
These insights bear importance for previous experiments \cite{Crook2000a,Burke2010,Ferry2011,Aoki2012} on scanning gate imaging of open quantum dots. Our results will lead to educated designs of future scanning gate experiments on cavities. Guiding branches with shallow potentials promises experiments in the realm of mesoscopic physics, in which caustics are controlled by external voltages. Our results raise the interesting theoretical question, which information about the disorder potential can be extracted from measurements of the branch pattern.

\begin{acknowledgments}
	
We thank Beat Br{\"a}m, Dietmar Weinmann, Rodolfo Jalabert, Guillaume Weick, and Boris Brun for helpful discussions. We acknowledge financial support from the Swiss National Science Foundation, the NCCR "Quantum science and Technology" and ETH Z{\"u}rich.

\end{acknowledgments}

\bibliography{branchesBib}

\end{document}